\documentclass[11pt,a4paper]{article}

\usepackage[a4paper,margin=3cm]{geometry}
\usepackage{lmodern}
\usepackage[T1]{fontenc}
\usepackage[utf8]{inputenc}
\usepackage{graphicx}
\usepackage{amsmath}
\usepackage{booktabs}
\usepackage{tabularx}
\usepackage{url}
\usepackage{natbib}
\usepackage[hidelinks]{hyperref}

\DeclareUnicodeCharacter{03B1}{\ensuremath{\alpha}}
\DeclareUnicodeCharacter{03B2}{\ensuremath{\beta}}
\DeclareUnicodeCharacter{03BC}{\ensuremath{\mu}}
\DeclareUnicodeCharacter{03C3}{\ensuremath{\sigma}}
\DeclareUnicodeCharacter{2212}{\ensuremath{-}}

\makeatletter
\def\maxwidth{\ifdim\Gin@nat@width>\linewidth\linewidth\else\Gin@nat@width\fi}
\makeatother
\setkeys{Gin}{width=\maxwidth,keepaspectratio}

\mathchardef\UrlBreakPenalty=100
\mathchardef\UrlBigBreakPenalty=100
\newcommand{\Polisp}{\texorpdfstring{Polis\textsuperscript{+}}{Polis+}}
\newcommand{\Polispp}{\texorpdfstring{Polis\textsuperscript{++}}{Polis++}}

\graphicspath{{../../figures/}{figures/}}

\newif\iffinaldraft\finaldraftfalse
\InputIfFileExists{finaldraft.flag}{\finaldrafttrue}{}
\iffinaldraft

\else

  \usepackage{xcolor}
\fi

\newcommand{\prov}[1]{#1}

\title{It Takes Three to Converse:\\[2pt]
       \large Empirical Observations on How the Developer, the Convener and the
       Participant Shaped 119 Polis Conversations}
\author{Lodewijk Gelauff\thanks{Research Scientist, Generation Lab; Research Affiliate, Stanford Deliberative Democracy Lab. \texttt{lgelauff@gmail.com}. This work was carried out in the author's personal capacity; neither organisation had any role in it.}}
\date{August 2026}

\begin{document}
\typeout{ARXIVDIMS textwidth=\the\textwidth}
\typeout{ARXIVDIMS textheight=\the\textheight}
\typeout{ARXIVDIMS oddsidemargin=\the\oddsidemargin}
\typeout{ARXIVDIMS topmargin=\the\topmargin}
\typeout{ARXIVDIMS paperwidth=\the\paperwidth}
\typeout{ARXIVDIMS paperheight=\the\paperheight}
\typeout{ARXIVDIMS headheight=\the\headheight}
\typeout{ARXIVDIMS headsep=\the\headsep}
\maketitle

\begin{abstract}

Polis is a popular democratic innovation tool that allows asynchronous citizen engagement through atomic statements: short statements that together describe a complex question, inviting the citizen to vote Agree or Disagree on each. This paper uses 119 conversations with 100 or more participants and an extensive data export, drawn from a wider set of 271 collected processes. 

The paper asks what determines the output of such a process. Three parties shape the result. The developer of the platform has made important design choices that restrict the outcome: the number of groups the platform is able to report (restricted to 2--5) and which statements are prioritized. 

The convener defines the assignment: the initial statements that set the tone, the policy that accepts or rejects new statements and who can be invited. Finally, the participant works within these boundaries. With access to less than half of the generated statements, they end up responding to more statements when their conversation seems to have an achievable number of statements to complete, than when they are presented with more statements. 

Due to choices such as warm path clustering, the exported resulting clustering cannot be reproduced based on the voting data. Conveners may want to re-analyse their own conversations once the process is closed, to consider the data in its entirety, and make their own analysis priorities explicit. 

\end{abstract}

\section{Introduction}
Citizen engagement is adopted increasingly broadly, and many digital tools are available to policy-makers to do so \citep{deseriis2023reducing, frenkiel2025boosting}.
A popular family within these digital tools is the Polis-family: asynchronous conversations where citizens respond to atomic statements that jointly describe a complex problem. Citizens then participate in this conversation by responding to the short statements with Agree, Disagree, Pass. At the end of the process, the citizen and the convener gain insight into how these multi-dimensional opinions cluster together, where the participants agree, and where they do not \citep{small2021polis, grossi2026computational}.

These platforms provide policy makers with a wide range of configurations, implicit or explicit. For example, the convener can seed the conversation with statements that describe the topic, and participants can then submit additional statements~\citep{small2021polis}. 
Whether the convener sets a strict or a permissive moderation policy may determine how rapidly the number of available statements grows, and what the experience of the participants looks like (see Section~\ref{sec:moderation-mechanisms-available}).

In the past years, the platform and its offshoots have been popular with a range of local and national governments in Finland \citep{sitra2023polis}, Germany \citep{heise2018aufstehen}, the Netherlands \citep{zuidholland_polis}, the United Kingdom \citep{demos2020polis}, the United States \citep{compdem_kentucky} and famously Taiwan \citep{hsiao2018vtaiwan, yang2025bridging}. Besides governmental conveners, also non-governmental organizations including Anthropic \citep{huang2024collective} and Demos/Open Rights Group \citep{demos2020polis} are known to have used these conversations in their stakeholder engagement. 

I was able to track down mentions of \prov{293} such conversations, and obtained usable data for \prov{271}. This paper describes a dataset with extensive opinion data from 119 conversations and analyzes their dynamics. It provides new insights into the effects of the platform's design choices, into the choices conveners made in this data, and into how participants behaved. No index of Polis conversations exists, so no collection can be exhaustible and ours is not either. What it can offer is scale and documented implementation: 119 conversations with complete vote records. 

With these records I demonstrate the effects of some of the design choices in the software, which may be of value to both platform designers and conveners, such as the effect of having an exhaustible stack of statements. 

On the platform, short statements are presented to participants and they are asked whether they agree or disagree. Besides a seed set that is provided by the convener, participants can also submit additional statements themselves. While this may seem at first a simple interaction, the result is a landscape defined by clusters of participants and their opinions that invites further examination. It is open source software and therefore it may not be surprising that there is a number other platforms inspired by this approach, that use slightly different implementations. I refer to the immediate set of forks and reimplementations that aim to stay as close as possible to Polis as \Polisp{} and to the broader family as \Polispp{}.

Another consequence of the fact that Polis is open source is that the software can be self-hosted by the convener. This means there is no public index of all conversations, and the overviews I was able to track down were privately maintained rather than published.  Thanks to the open approach to data in this process, the output is typically not just the resulting clusters, but also the underlying data that describes the state of the analysis engine at the moment of export. 

The opinion landscape that is described by the clusters at the end of the conversation is routinely narrated as a finding about a public: facilitator reports, press, and the research literature alike lead with the number of clusters that was the outcome \citep{hsiao2018vtaiwan, huang2024collective, small2023opportunities, demos2020polis, paice2022integrating}. 
I show that the report of the conversation with its clusters represents that same state of the engine: it is not necessarily the only or the best representation of the conversation as a whole. I argue that a convener should run their own analysis and draw their own conclusions after the process has completed, with the added benefit of being able to clean the data thoroughly first. 

For the benefit of the reader, I will briefly describe how the paper is organized. Section~\ref{sec:background} introduces the platform, terminology and the wider family of platforms it belongs to, together with the existing relevant literature. Section~\ref{sec:data} describes the data collection and cleaning, and what is included. 

In the next three sections, the data is analysed. Section~\ref{sec:developer-choices} describes some of the design choices of the platform, and how they affect the process and its results: how groupings are produced, which statements are routed to participants and how the author can submit alternative statements. Section~\ref{sec:convener-choices} turns to how organisers used the platform: seeding of statements and moderation of the conversation. Section~\ref{sec:participant-choices} describes the patterns that emerge from their voting behavior. Section~\ref{sec:outputs} then describes how the opinion landscape is reported in clusters. %
Finally, Section~\ref{sec:limitations} discusses the limitations and what the implications might be for practitioners, and Section~\ref{sec:conclusion} summarizes the generalizable findings.

\section{Background}
\label{sec:background}
Polis is one of the systems that can be used to better understand what an opinion landscape looks like~\citep{grossi2026computational}.
The Polis system used the recommender systems known for dating apps and gig platforms to create a platform for crowdsourced 'wiki' surveys \citep{salganik2015wiki, small2021polis, bass2019crowdsourcing}. The first contributions stem from 2012, and got better known after its implementation  by the Sunflower Movement in Taiwan in 2014-2015. 
Polis has evolved over the past years, as have its spin-offs. For example, both Agora and Voxit no longer use the Polis mathworker as per August 2026, and several of the findings may therefore no longer apply to them.

This section provides the terminology that grounds the rest of the paper, and points out some unexpected implementation details or `quirks' that may matter for the implementation or interpretation of the results.

Well-known examples of Polis deployments include the vTaiwan platform, the Austrian Klimarat, and the German Aufstehen movement.
The implementation by vTaiwan is perhaps the most referred citation in popular literature. Polis was used in 2018 as part of a larger process that involved multiple democratic innovation tools in order to facilitate a public consultation on how to regulate Uber in Taiwan \citep{hsiao2018vtaiwan}.
The Austrian \emph{Klimatrat} refers to a 2022 national climate assembly over six weekends, running alongside five Polis conversations as input for those assemblies. These egaged more than 5{,}000 people, and is particularly interesting because of the repetitive element. Both a facilitator account and an independent academic evaluation were published \citep{paice2022integrating}.
The implementation by the Aufstehen movement is a fork of the upstream Polis software, and it is the largest deployment of the platform in this dataset, with 33,547 participants and over 2 million opinions expressed. Notably, 26\% of its responses were in the first 24 hours. \emph{Aufstehen} (``Stand Up'') was a 2018 left-wing collection movement in Germany that asked early supporters what its movement should stand for \citep{heise2018aufstehen}.

The platform has been used in a range of different contexts, including citizens' assemblies, party platform-building, and government consultation.

\subsection{How Polis works}
\label{sec:polis-works}
Overall, Polis is a platform for asynchronous consultation. The \textbf{convener} of the process has control over the settings of the \textbf{conversation}. They contextualize the conversation, select \textbf{seeding statements} and \textbf{meta-statements}, decide on authentication or participation limitations, set a priority policy for statements, moderation policy (strict or not) and will be able to remove statements, approve or reject moderated statements, set data transparency. Maybe more importantly, they also effectively decide who gets to know about the conversation: how is it advertised and when are reminders sent.

Consider a toy conversation that has as purpose to determine whether cats or dogs make the better pet. Participants are recruited and provided with a link to the \textbf{conversation} on the conveners' server or that of a host such as pol.is. A participant may be presented with an authentication option, after which they can respond to a number of \textbf{statements} such as `Dogs are more loyal' and `Cats require less care'. These are presented one by one on a \textbf{card}. The first cards will be presenting meta-statements (`I currently own a cat') that can serve as demographics to interpret the resulting clusters. Based on the configuration, statements are presented sequentially, prioritised when seeded or randomly drawn with weights determined by a \textbf{priority function}.

\begin{figure}[htbp]
\centering
\includegraphics[width=0.82\textwidth]{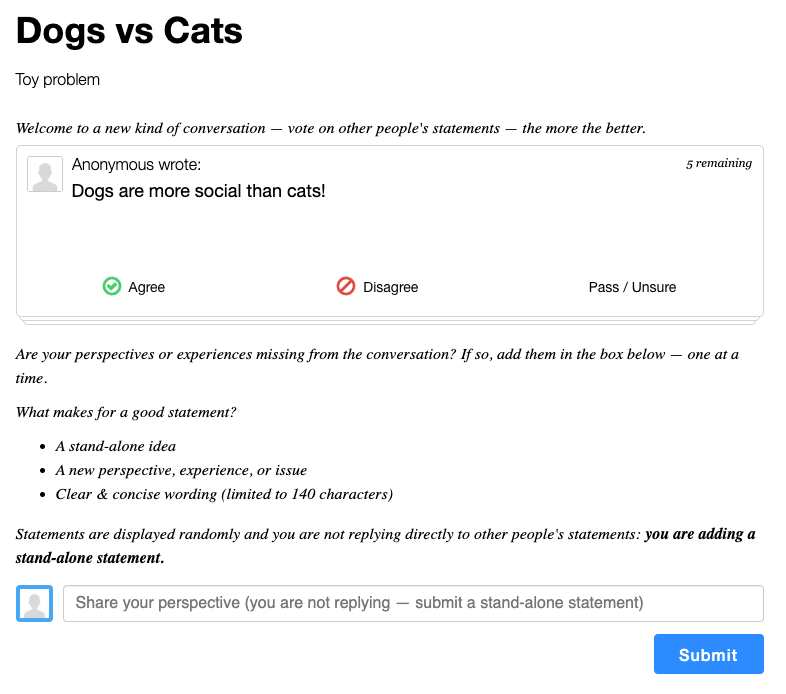}%
\caption{The participant's view of a statement, from an example conversation of my own. One statement is shown at a time on a card, with the three responses the platform allows --- agree, disagree
and pass --- and a count of how many statements remain. Below the card, a participant may submit a
stand-alone statement of their own rather than reply to the one shown.}
\label{fig:ui-card}
\end{figure}

To each statement, a participant can express that they agree, disagree or they can pass. Alternatively, they can submit an alternative statement such as `rabbits are undervalued'. Depending on the moderation policy, this is then available to other participants either immediately or upon approval from the convener. The participant responds to these statements one by one, and can leave at any time. The platform will then cluster the participants based on their responses, and present them with a summary of the clusters. The convener can then use this information to inform their decision-making process.

\subsection{The \Polisp{} and \Polispp{} family}
\label{sec:family}
Polis has inspired a number of similar platforms that operate in comparable ways. I refer to these as \Polisp{} and \Polispp{} platforms. The former are platforms that use the same software as Polis, but may have different interfaces or branding. The latter are platforms that have reimplemented the method in other code, but still operate in a similar way. Tables~\ref{tab:polis-the-same-software-or-a}, \ref{tab:polis-reimplementations} and~\ref{tab:polis-interfaces} summarise these platforms: the same software under another name, reimplementations of the method, and interfaces over an unmodified engine.

\begin{table}[htbp]
\centering
\footnotesize
\caption{\Polisp{}: the same software, or a fork of it.}
\label{tab:polis-the-same-software-or-a}
\begin{tabularx}{\textwidth}{@{}>{\raggedright\arraybackslash}X>{\raggedright\arraybackslash}X>{\raggedright\arraybackslash}X>{\raggedright\arraybackslash}X>{\raggedright\arraybackslash}X@{}}
\toprule
\textbf{Platform} & \textbf{Code repository} & \textbf{Website} & \textbf{Timespan} & \textbf{Example deployments} \\
\midrule
Polis (upstream) & \path{compdemocracy/polis} & \path{compdemocracy.org} & 2012-- & vTaiwan, \emph{Aufstehen}, Klimarat Österreich \\
PolisNL & frontend: Partici.app / ParticiAPI & \path{polisnl.org} & 2021--2025 & 10 published conversations, 118--1,213 participants each \\
Voxit (Sitra / DigiFinland)\footnote{Voxit ran the Polis engine over the period the Finnish conversations in this corpus were collected. It was rebuilt in 2026 on a different math implementation, and appears in Table~\ref{tab:polis-reimplementations} for that period.} & \path{gitlab.com/voxit} (18 projects) & --- & --2025 & Finnish national deployment \\
DEMDIS & \path{Demdis/polis} (detached copy) & --- & --- & Slovak \\
DfE whitelabel & \path{DFE-Digital/polis-whitelabel} & --- & 2025--2026 & UK Department for Education \\
polis.tw & \path{PDIS/polis2023} & \path{polis.tw} & 2023-- & Taiwan \\
Kennislink & \path{Simon-Dirks/polis} & --- & --- & --- \\
Crowd Wisdom Project & \path{NewRedo/polis-fork} & --- & --- & --- \\
PolisOrbis (Copernicani) & \path{gitlab.com/copernicani/polisorbis/codebase} & \path{orbis-project.eu} & 2024--2026 & Italian EU project \\
HIERR & \path{CodeWithAloha/HIERR} & --- & 2025 & Hawai'i Economic Recovery \& Resilience, State OPSD \\
\bottomrule
\end{tabularx}
\end{table}

\begin{table}[htbp]
\centering
\footnotesize
\caption{\Polispp{}: platforms that reimplement the method in other code. A convener runs the
conversation here, but not on the engine analysed in Section~\ref{sec:developer-choices}.}
\label{tab:polis-reimplementations}
\begin{tabularx}{\textwidth}{@{}>{\raggedright\arraybackslash}X>{\raggedright\arraybackslash}X>{\raggedright\arraybackslash}X>{\raggedright\arraybackslash}X>{\raggedright\arraybackslash}X@{}}
\toprule
\textbf{Platform} & \textbf{Code repository} & \textbf{Website} & \textbf{Timespan} & \textbf{Example deployments} \\
\midrule
Agora Citizen Network & \path{zkorum/agora} & --- & 2023-- & --- \\
EJ Platform (``Empurrando Juntas'') & \path{ejplatform/ej-conversations}, \path{ej-math} & --- & 2018--2022 & Brazil \\
Nexus (was MindMeld) & \path{sofvanh/Nexus} & --- & 2025-- & --- \\
Voxit, from 2026 & \path{gitlab.com/voxit} & --- & 2026-- & Finnish national deployment \\
\bottomrule
\end{tabularx}
\end{table}

\begin{table}[htbp]
\centering
\footnotesize
\caption{Interfaces over an unmodified Polis engine. These change what a participant sees, not
what is computed.}
\label{tab:polis-interfaces}
\begin{tabularx}{\textwidth}{@{}>{\raggedright\arraybackslash}X>{\raggedright\arraybackslash}X>{\raggedright\arraybackslash}X>{\raggedright\arraybackslash}X>{\raggedright\arraybackslash}X@{}}
\toprule
\textbf{Platform} & \textbf{Code repository} & \textbf{Website} & \textbf{Timespan} & \textbf{Example deployments} \\
\midrule
Metropolis (Fil Poll) & \path{canvasxyz/metropolis} & \path{poll.fil.org} & 2024--2025 & Filecoin governance \\
Partici.app / ParticiAPI & --- & --- & 2021--2025 & PolisNL \\
Polis Japan & \path{PolisJAPAN/PolisJAPAN} & --- & 2026 & frontend on the \path{pol.is} server \\
\bottomrule
\end{tabularx}
\end{table}

Two further projects reimplement only the mathematics. Red Dwarf, a Python implementation of the Polis pipeline, now computes the opinion groups for both Agora and Voxit; the Brazilian Pentano did the same in 2017, eight years earlier. Neither is a platform a convener would choose --- they are the layer underneath one --- but they are the reason a \Polispp{} platform can produce a Polis-shaped result from code that upstream never wrote. Nor do these platforms only reimplement: Agora does not impose an order on the statements a participant sees, letting them choose their own, and offers methods beyond the one Polis uses.

\subsection{Previous work}
\label{sec:previous-work}
There are a few papers that describe the platform Polis itself in more detail. 
\citet{small2021polis} describe how the platform developers approached the scaling challenges, the mechanics behind their clustering algorithm, and experiments with implementing large language models. %
In the paper the authors describe the basic principles behind the system and how the resulting landscape of clusters is calculated through an iterative expectation-maximisation. The authors argue that the chosen approach is justified because the vote matrix is by construction sparse because of the option to submit additional statements. 
Polis uses a math engine based on cycles of power-iteration PCA \cite{roweis1998em}. Polis uses the principal component eigenvectors from the previous cycle, and introduced the application of all other steps needed to arrive at a new clustering in the same cycle. This includes the application of moderation decisions, picking the desired number of clusters and subclusters subject to a smoothing function. There is some precedent in evolutionary clustering \citet{chakrabarti2006evolutionary, chi2007evolutionary}. 

The project also maintains public documentation of the method \citep{compdem_algorithms}. The mechanisms this paper examines are therefore described upstream first, and not discovered here, although some experimentation and source code archaeology was sometimes needed to bridge the gaps.

The method is partially inspired by the older wiki survey \citep{salganik2015wiki}, which described three principles for open-ended data collection: it should be greedy, collaborative and adaptive to responses.
The shifting set of statements is an essential part of the design. In more recent work from the same group, the use of large language models alongside the traditional system is considered for facilitation, moderation and summarisation \citep{small2023opportunities}. 

The atomic statement approach has been taken up beyond sketching an opinion landscape: \citet{huang2024collective} used Polis with a representative sample of roughly a thousand US adults to draft a foundational document for language models.

Reports from individual deployments are frequently published, although often outside peer-reviewed venues. 
The vTaiwan consultation on ride-hailing platforms was reported
\citep{hsiao2018vtaiwan} and has been revisited in a recent field study
\citep{yang2025bridging}. 
In the United Kingdom, a consultation on campaign data resulted in a
commissioned report \citep{demos2020polis}, and the Austrian Klimarat was reported both by a facilitator account and through an academic evaluation \citep{paice2022integrating, buzogany2022evaluation}. 

Publications comparing different digital methods are rare: \citet{frenkiel2025boosting} compares the use of Polis and Decidim in a student assembly, and \citet{deseriis2023reducing} contrasts six decision-making platforms. Those studies compare how the process would run under different conditions. Here I pose a different question: what the available datasets from those consultations actually contain, and what can be learned from them about the process. 

Polis does have some similarities to bridging models that also make use of or describe the opinion landscape. Community Notes does not pose any questions, but scores contributions on a latent space \citep{wojcik2022birdwatch}. Framing bridging systems generally, \citet{ovadya2023bridging} name Polis as the discrete case.

A few publications engage with the resulting data from Polis conversations. In a randomised controlled trial, \citet{venkat2024efficacy} presented the summary from Polis against one from a computational-social-choice and an abstract-argumentation algorithm on a nationally representative sample, and reported that the social-choice method left participants feeling better represented than the Polis one. 
Polis was also analyzed in a perspective on deliberative digital media \citep{pentland2024toward}. In both cases the output was approached as such, where this analysis approaches it as a way to study its provenance. I did not identify a study of the outputs at scale at the time of writing.

\section{Data}
\label{sec:data}

\subsection{Collection and cleaning}
\label{sec:collection-cleaning}
Polis conveners have the option to indicate their data to be open or not, and to mint reports. Once a report is minted, the opinion data is available in anonymous form for download once the location of the website is known. For this data collection, I searched online for mentions of these Polis conversations and downloaded these datasets as CSV files. By its nature, this is the snapshot as available at the time of download.

I searched online iteratively for public mentions of Polis conversations and for websites that had a signature similar to that of a Polis conversation. I also included some digital archives in this search, to reach venues that were not using the same terminology. After each identified result, I updated the queries. 
I identified \prov{329} raw export directories, each holding a \path{votes.csv} file. The location of such an export was recorded and archived at the Internet Archive to establish provenance if possible. Deduplication of these removed \prov{56} entries. Three more were excluded: one that was negligible (a single statement and seven votes) and two that were clear demonstrations. One further conversation was \emph{added} rather than excluded --- it was published in a spreadsheet rather than the standard export, and was recovered through a separate adapter. This left \prov{271} conversations with \prov{8{,}628{,}379} votes.

Next, I removed 2 conversations because of vote-integrity concerns, 50 because of the absence of a single clean event window and 94 because of having fewer than 100 voters; a further 6 were quarantined. This left \textbf{\prov{119} conversations} and \prov{6{,}883{,}979} votes. 

Based on the voting patterns I established the most likely voting window: the vast majority held clear single peaks, but several pages were kept open after the end of the obvious time window. In those cases I used the creation of new statements and when that stopped (i.e.\ when moderation activity ended), or a clear drop in activity, to establish what the voting window was. Throughout I erred on the side of safe margins, including only data that was clearly part of the original conversation and its recruited participants. 

\subsection{Data set description}
\label{sec:data-set-description}
Most of this paper uses the 119 conversations that resulted from the collection process described above. For some metrics I report the 271 instead, where that is more relevant. 

Unless stated otherwise, every figure in this paper is over these 119 conversations. Where a figure is reported over another set, that is stated explicitly. Four bases recur and they are not interchangeable; Table~\ref{tab:bases} sets them out. A fifth distinction matters inside the 119: the cohort carries \prov{6{,}883{,}979} votes in total, while the analyses run on the \prov{6{,}540{,}474} that fall inside a conversation's established voting window.
\begin{table}[htbp]
\centering
\footnotesize
\caption{The four bases used in this paper. They are not interchangeable, and
they are not successive filters of one another: the census includes leads for
which no data was ever obtained, and so is not a superset of the corpus.}
\label{tab:bases}
\begin{tabular}{@{}rll@{}}
\toprule
\textbf{n} & \textbf{Base} & \textbf{What it counts} \\
\midrule
329 & raw export directories & directories on disk containing a \path{votes.csv} \\
293 & the census             & catalogued conversations, including leads we never obtained \\
271 & the cleaned corpus     & deduplicated, canonicalised and flagged \\
119 & the analysis cohort    & single-event, clean timeline, $\geq$100 voters, not quarantined \\
\bottomrule
\end{tabular}
\end{table}

\begin{figure}[htbp]\centering
\includegraphics{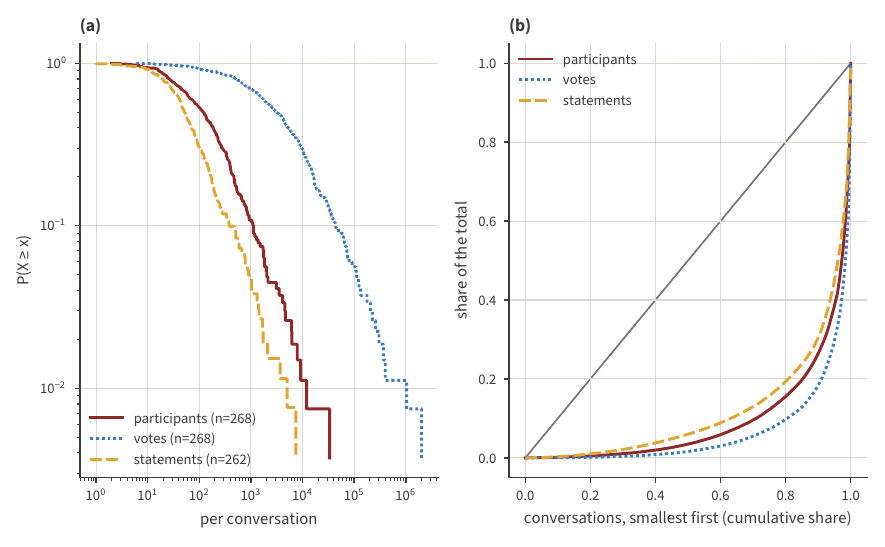}
\caption{How unevenly the corpus is distributed across conversations.}
\label{fig:corpus-skew}
\end{figure}

The votes of the 119 data sets are distributed between 2015-01-11 and 2026-05-24, with conversation sizes ranging from \prov{45} to \prov{33{,}421} (mean \prov{955}, median \prov{332}) voter IDs and from \prov{921} to \prov{2{,}013{,}990} (mean \prov{54{,}962}, median \prov{10{,}449}) votes. Both distributions are heavily skewed; see Figure~\ref{fig:corpus-skew}.
\begin{table}[htbp]
\centering
\footnotesize
\caption{The analysed corpus, per conversation. Medians with interquartile range, over the 119 conversations in the analysis cohort. Spans are of the \emph{analysis window}, not of the conversation's full lifetime.}
\label{tab:corpus-shape}
\begin{tabular}{lrrrr}
\toprule
 & \textbf{Median} & \textbf{Q1} & \textbf{Q3} & \textbf{Max} \\
\midrule
Participants & 332 & 176 & 552 & 33,421 \\
Statements & 128 & 73 & 262 & 7,378 \\
Votes & 10,449 & 4,874 & 25,172 & 2,013,990 \\
Votes per participant & 32.5 & 25.4 & 51.3 & 131.4 \\
Votes per statement & 78.9 & 51.4 & 120.1 & 3,284.8 \\
Span (days) & 11.1 & 5.6 & 18.2 & 51.9 \\
\midrule
\multicolumn{5}{l}{\emph{Totals:} 6,540,474 votes, 43,542 statements, 113,628 participant--conversation pairs} \\
\bottomrule
\end{tabular}
\end{table}

Table~\ref{tab:corpus-shape} gives the shape of the corpus per conversation. Every quantity is long-tailed in the same direction: the median conversation is small and the mean is pulled well above it by a handful of very large events.

\begin{figure}[htbp]\centering
\includegraphics{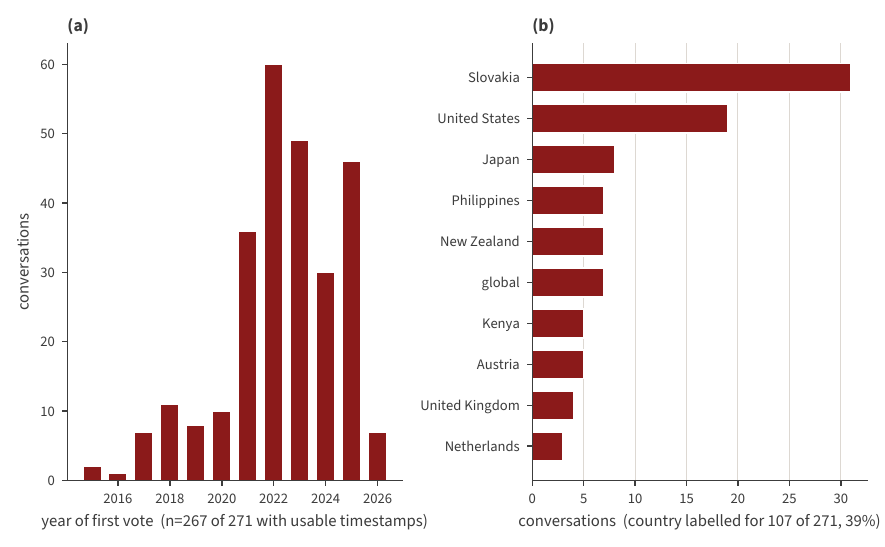}
\caption{Conversations by year of first vote, and the ten commonest countries.}
\label{fig:corpus-context}
\end{figure}

For the originally identified 271 datasets, country and language are labelled for \prov{107} (\prov{39}\%) and sector for \prov{80} (\prov{30}\%).

\begin{table}[htbp]
\centering
\footnotesize
\caption{Countries, over the 82 of 119 conversations for which a country could be established. \textbf{The unlabelled remainder is not a random sample:} an export found through a national algorithm register carries a country, one found through a bare report link usually does not, so this table describes how conversations were found as much as where they happened.}
\label{tab:corpus-countries}
\begin{tabular}{lrr}
\toprule
\textbf{Country} & \textbf{Conversations} & \textbf{Share of labelled} \\
\midrule
Slovakia & 22 & 27\% \\
United States & 12 & 15\% \\
New Zealand & 7 & 9\% \\
Philippines & 7 & 9\% \\
Japan & 6 & 7\% \\
global & 5 & 6\% \\
Austria & 5 & 6\% \\
Kenya & 4 & 5\% \\
United Kingdom & 3 & 4\% \\
Netherlands & 3 & 4\% \\
Other (6 countries) & 8 & 10\% \\
\midrule
Unlabelled & 37 & --- \\
\bottomrule
\end{tabular}
\end{table}

The distribution below describes where a country could be \emph{established} as much as where conversations happened: an export reached through a national algorithm register carries one, while an export reached through a bare report link usually does not.
Across the \prov{271} collected conversations, \prov{256} carried an `operator string' identifying the convener, from \prov{139} named organisations; the largest, DEMDIS in Slovakia, held \prov{31}. Within the \prov{119} analysed here the concentration is similar: \prov{58} operators, of which DEMDIS holds \prov{22}.

\begin{figure}[htbp]\centering
\includegraphics{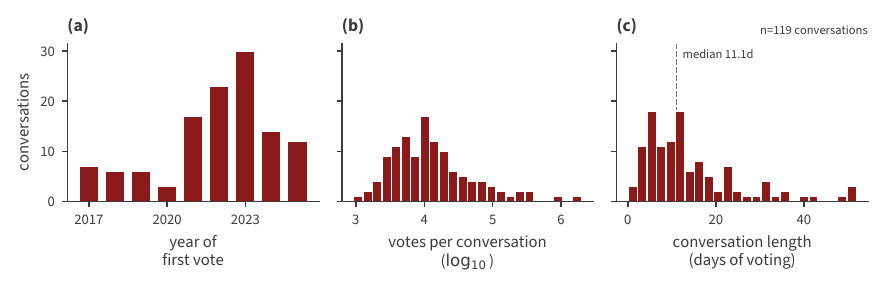}
\caption{The corpus in two views: when conversations started, how many votes each drew, and how long its analysis window ran. All three axes carry their base.}
\label{fig:corpus-windows}
\end{figure}

All \prov{119} conversations carry a terminal moderation policy; \prov{112} a group count, \prov{22} an end date and \prov{9} a seed-priority setting. The established timespan for the conversations ranges from under a day to \prov{52} days (median \prov{11}), and the distribution is displayed in Figure~\ref{fig:corpus-windows}.
\begin{table}[htbp]
\centering
\footnotesize
\caption{How long voting ran, over the 119 conversations in the analysis cohort. This is the span of votes \emph{within} each conversation's established voting window (Section~\ref{sec:collection-cleaning}), so the 52-day maximum is a property of the measure rather than of the conversations --- a page is often left reachable long after its event. The untrimmed span is not a better answer: stray later votes put 24 of these conversations past 1{,}000 days, and the largest past three centuries.}
\label{tab:corpus-windows}
\begin{tabular}{lrr}
\toprule
\textbf{Window length} & \textbf{Conversations} & \textbf{Share} \\
\midrule
Under 24 hours & 2 & 1.7\% \\
1--7 days & 31 & 26.1\% \\
7--30 days & 72 & 60.5\% \\
Over 30 days & 13 & 10.9\% \\
\midrule
Median & \multicolumn{2}{r}{11.1 days} \\
First vote in corpus & \multicolumn{2}{r}{2017-06-14} \\
Last vote in corpus & \multicolumn{2}{r}{2025-12-05} \\
\bottomrule
\end{tabular}
\end{table}

Most conversations are only open for a short amount of time, which demonstrates the wide range of use cases that Polis serves. Table~\ref{tab:corpus-windows} describes the participation window of these conversations, as estimated from the data. Note that many processes were still accessible after this point. 

Most statements draw more agreement than disagreement: \prov{81.7}\% of the \prov{21{,}432}
statements that received at least ten votes from someone other than their author. The share is
insensitive to where the threshold is put, ranging from \prov{80.2}\% at one vote to \prov{83.6}\%
at fifty.

\subsection*{How the figures in this paper were checked}
Every reported interval is a bootstrap that resamples whole conversations rather than individual
observations, because participants nest inside conversations; where an operator-level interval is
also available it is reported alongside, and it is the wider of the two. Every randomised step uses
a fixed project seed, and re-runs are byte-identical. Where a rate depends on a threshold, it is
reported at more than one threshold. Where many windows or buckets are tested at once, $p$-values
are corrected for multiple comparisons before any pattern is described.

\subsection{Export data}
\label{sec:export-record}
The export data can contain the following files:
\begin{itemize}
  \item \path{summary.csv} --- one key--value row per field: topic, url, voters, voters-in-conv, commenters, comments, groups, conversation-description
  \item \path{comments.csv} --- one row per statement, with its author, timestamp, moderation state and agree/disagree/pass tallies
  \item \path{votes.csv} --- one row per vote: timestamp, voter, statement, value
  \item \path{participant-votes.csv} --- the vote matrix, participants as rows and statements as columns, with the assigned group
  \item \path{participant-importance.csv}, \path{comment-groups.csv}, \path{comment-clusters.csv} --- derived per-group outputs at the time of export
\end{itemize}

Based on the voting and statement submission patterns, I was able to identify conversations that used specific moderation settings. 

Because new statements without votes are given a high priority by the routing algorithm, the statement can be assumed to have been approved for voting at roughly the moment the first vote arrives from someone other than its author. This allows the `lifespan' of a statement to be estimated even when the approval and rejection are not explicitly recorded. I labelled this in the dataset, assuming at most a single approval time and at most a single rejection time. Unknown is explicitly permitted as a possible outcome. 

\begin{itemize}
  \item Demonstrated visibility. A statement counts as available to a participant once someone other than its author has voted on it. Using ``created before t'' instead overstates the unvoted pool, because it counts statements that were rejected, never shown, or never approved.
  \item Estimated end of visibility. For a rejected statement, the time of its last received vote is taken as the end of its visibility: a hidden statement can receive no further votes, so the last vote is a lower bound on the moment of removal. Statements rejected before anyone voted get no estimate, and approved statements need none --- they remain visible to the end.
  \item \textbf{Sessions} for a participant follow a 1-hour inactivity rule; they are not observed sessions
\end{itemize}

\subsection{Two kinds of conversation}
\label{sec:distributions-two-kinds}
The number of votes varies widely between conversations, reaching millions at the largest. The median conversation in this corpus drew \prov{118} participants and the median participant voted \prov{25.5} times. See also Table~\ref{tab:corpus-shape}.
The number of submitted statements likewise ranges all the way to \prov{7{,}400}, against a median of \prov{132}. However, this is not the same as the number of statements that was actually observed by the participants. 

Evaluating what portion of their votable pool the median participant completed in each conversation reveals a clear split. On one hand there is a large number of people who voted on every statement they could possibly vote on; on the other, a peak of people who voted on only a small share of the statements available to them. In absolute terms, however --- how many statements they voted on --- the two kinds of conversation are hardly distinguishable. 
\begin{figure}[htbp]\centering
\includegraphics{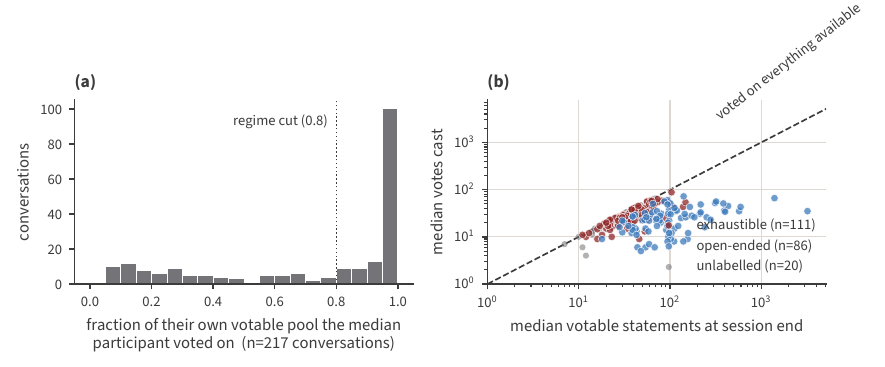}
\caption{What share of the statements available to them participants responded to, by kind of
conversation. The two regimes differ far more in this share than in the number of votes cast.}
\label{fig:bimodality}
\end{figure}

These are two types of conversations that are being organized. On one hand there are conversations where a small number of seeded statements is made available and no other statements are being approved for consideration. On the other hand there are more open-ended conversations, where each statement that is proposed can also be voted on. Finally there are conversations in the middle, where a smaller number of statements is approved for voting. 

These types are easy to recognise once looked for, and may explain a number of different data patterns.

\section{The developer's choices}
\label{sec:developer-choices}
In this section I describe findings from this dataset that help understand the design choices in Polis and their effects. Other editions in \Polisp{}, or even different versions of Polis, may have made different choices, and behave differently. Not all of these are decisions as such. Some 'choices' might be coding decisions or even errors in the source code. That does not make the findings less relevant: it demonstrates how the design choices affect the outcomes.

\subsection{Routing and exposure}
\label{sec:routing-exposure}
From the exports it is possible to calculate the state the conversation should have been in, and therefore what the intended priority distribution would have looked like. Until January 2018 the priority of a statement was uniformly distributed. Between March 2025 and July 2026 the priority was also uniformly distributed due to a bug in the source code. This section focuses on the intended priority for 2018--2025 and from July 2026 onward.
The priority of a statement can be calculated based on a few formulas that can be derived from the source code:

\noindent Writing $A$, $D$ and $P$ for the number of agree, disagree and pass responses a statement has
received, $S$ for the total number of responses for it, and $E$ for its extremity, the smoothed rates and
the priority weight are\footnote{
  \texttt{compdemocracy/polis}, \path{math/src/polismath/math/conversation.clj}, lines 311--330 at commit \texttt{424dcae0}.
}
\begin{align*}
  p \;=\; \frac{P+1}{S+2}, \qquad a \;&=\; \frac{A+1}{S+2},\\[2pt]
  \mathrm{importance} \;&=\; (1-p)\,(E+1)\,a,\\[2pt]
  \mathrm{priority} \;&=\; \bigl[\,\mathrm{importance}\cdot(1 + 8\cdot 2^{-S/5})\,\bigr]^{2},
\end{align*}
with meta-statements held at a fixed $7^{2}=49$. The $+1/+2$ priors mean a fresh statement starts
neutral at $a=p=\tfrac12$, and that priority can never reach exactly zero, only become very small.
Extremity is the Euclidean distance of the statement from the origin in the two-dimensional projection, so the $(E+1)$ term rewards statements that separate the groups. 

These choices have some routing consequences:
\begin{itemize}
  \item The novelty bonus decays exponentially in votes received: the $1 + 8\cdot2^{-S/5}$ term starts at $9$ for an unvoted statement and falls towards $1$ --- $5$ after five votes, $3$ after ten, $2$ after fifteen. Priority squares it, though the net advantage to a new statement is considerably smaller than that squaring suggests.\footnote{
    Squared, the term spans $81\times$ between an unvoted statement and the value it approaches, with $9\times$ remaining after ten votes and $4\times$ after fifteen. The net advantage is smaller because $S$ enters the importance term as well: a fresh statement sits at $p=\tfrac12$, the largest pass penalty it will ever carry, and that penalty relaxes as votes arrive. A new statement outranks an evenly divided statement with twenty votes by roughly tenfold, not eightyfold.}
  \item If the first few participants pass on the statement, that may lower the priority drastically for a long time, and the statement may rarely be seen by others.  
  \item The priority formula is not symmetric in agrees and disagrees. Holding extremity constant, at the same 10 total votes per statement an all-agree statement's priority will outweigh an all-disagree's by 121x, and at 30 votes this is 961x.
\end{itemize}

Because in Polis it is not possible for a participant to choose what the next statement will be to vote on, it is possible to estimate how close the actual assigned probability was to the intended probability.

\subsection{Submitting a statement}
\label{sec:submitting-a-statement}
In Polis, the platform assumes that the person submitting a statement, would also support it. If that statement is seeded by the organizer, then this is represented with the ID of the organizer. The author never had the opportunity to explicitly agree or disagree with their own submitted statement.

For each participant-submitted statement I construct the author's voting history at submission and predict their stance using every earlier statement they voted on and correlating it with the target statement across the voters the two share. Any author-injected votes are removed from the data. A pair of statements must share at least \prov{10} voters to contribute. This indicates whether the voters tend to treat the two statements as alike. For this, I can also consider future votes, under the assumption of no interaction between the voters. The author's own votes on these earlier statements are carried through the correlation and accumulated. 

\prov{6{,}980} of the \prov{14{,}105} participant-submitted statements (\prov{49.5}\%) give a usable score. Of these, \prov{1{,}289}, or \prov{18.5}\%, score negative: a predicted disagreement recorded as an agreement. I use the next vote of the author on a non-submitted statement as control, only varying the target, resulting in \prov{3{,}725} submission-control pairs in \prov{167} conversations. 
The instrument predicts a paired
difference in predicted stance of $+\prov{0.078}$. A percentile bootstrap that resamples whole
conversations with \prov{2{,}000} draws gives a \prov{95}\% interval of $[+\prov{0.052}, +\prov{0.107}]$. Repeating the control at the second and third statements the author voted on gives $+\prov{0.066}$ and $+\prov{0.076}$, so the difference is not an artefact of the control statement sitting adjacent in time to the act of authoring.

\subsection{How Polis produces a grouping}
\label{sec:polis-produces-grouping}
As the process is asynchronous, the votes arrive continuously. In the background, a \textbf{math worker} computes the new state of the machine in \textbf{ticks}: it starts with a random state, and each tick it calculates from the old state \texttt{[S]} and new votes (if available). It will apply any moderation decisions (replacing affected participants or statements with 0-vectors), use principal component analysis (power-iteration EM PCA) to arrive at a 2-dimensional projection of the data. It then performs k-means clustering and group k-means for candidate K. A smoothing function picks also in each step the reported K \texttt{[S]}. 
It uses a \textbf{warm path} method: each tick uses the state that resulted from the previous tick to determine the projection axes, layers of centroids for k-means clustering and the K counter \texttt{[S]}~\citep{small2021polis,roweis1998em}.

Implications of the chosen implementation:

\begin{itemize}
  \item The value for K is limited to a window \texttt{min(5, 2 + floor(count/12))} at both the group and subgroup level.\footnote{source: compdemocracy/polis, math/src/polismath/math/conversation.clj}
  \item Moderation affects the system at the moment of implementation - rejected statements may still affect the probabilities until the next calculation where it is considered.
\end{itemize}

\medskip\noindent\textbf{What the system reports, and whether it can be checked.}
The number of clusters the system reports depends on all three of the above -- the warm path, the math engine, and the permitted range for $K$. Even in the best of circumstances, choosing the correct $K$ can be complicated in $K$-means clustering \citep{rousseeuw1987silhouettes,hamalainen2017comparison}. 

Table~\ref{tab:published-k} summarises the distribution of the number of clusters reported by Polis, including data submitted after the end of the time window. It also reports the number of clusters that I recomputed as what the engine would have arrived at holding all parameters fixed, and only replacing the warm path with a clustering of the final state (this was possible for 112 conversations). The disagreement is mostly upward (33 increases, 9 descreases), rather than fewer clusters. Note that this is not claiming that the recomputed values are any more correct than the reported values -- they are simply different. It would require a more in-depth dive than can be justified for this work, to establish which would be more appropriate for each of the conversations. 

\begin{table}[htbp]
\centering
\footnotesize
\caption{The number of opinion groups, as reported by Polis and as recomputed here. \emph{Reported} is the value carried in the export; \emph{recomputed} applies the engine's own candidate range to the same final vote matrix, without the warm path. The two agree for 70 of the 112 conversations where both are available; the recomputation is unavailable for the corpus's largest conversation, whose 33{,}422 participants exceed the grid's cap. The engine's candidate range begins at two, so the selection cannot return a single group --- though an export can nonetheless carry one (see Sec.~\ref{sec:outputs}).}
\label{tab:published-k}
\begin{tabular}{lrrrr}
\toprule
\textbf{Groups} & \multicolumn{2}{c}{\textbf{Reported (114)}} & \multicolumn{2}{c}{\textbf{Recomputed (112)}} \\
\cmidrule(lr){2-3}\cmidrule(lr){4-5}
 & n & share & n & share \\
\midrule
2 & 85 & 74.6\% & 70 & 62.5\% \\
3 & 20 & 17.5\% & 15 & 13.4\% \\
4 & 6 & 5.3\% & 15 & 13.4\% \\
5 & 3 & 2.6\% & 12 & 10.7\% \\
\bottomrule
\end{tabular}
\end{table}

The grouping is invariant to the engine's own random starting point for the first round of PCA: six independent draws across \prov{180} conversations reproduce byte-identically in \prov{97.8}\% of cases. Invariance to the seed is not the same as reproducibility from the export, and the two should not be read as one result.

\section{The convener's choices}
\label{sec:convener-choices}
The settings are often not included as part of the available dataset, and when they are they only represent a snapshot of the final state.
However, based on the available data and on how participants interact with the conversation, some of the decisions that moderators made can be inferred. 

Once inferred, the effect of those settings on the participants can be observed. This gives some insight into how the moderation and organization of a conversation happen in practice rather than in theory. In this section I discuss the choices I observe conveners making, and the consequences of those choices for the conversation and its participants. 

\subsection{Moderation mechanisms available}
\label{sec:moderation-mechanisms-available}
Moderation functionality is advertised as optional by the platform designers, and instead framed as an efficiency tool. 

\begin{itemize}
  \item Seeding statements: the convener can add statements to the conversation before it opens, and can add additional statements mid-conversation. 
  \item The convener can set a priority for seeded statements, so that they are shown to participants first.
  \item The moderation scheme is a single toggle: strict moderation on/off in the conversation's configuration screen. Strict moderation is a whitelist: no comment is shown to participants unless the moderator has accepted it into the conversation. Permissive moderation is a blacklist: every comment is shown as soon as it is submitted, and can be moderated out afterwards.
  \item Three per-comment actions: approve, reject, or leave unmoderated. The third is a real state, not the absence of a decision.
\end{itemize}

The visibility of statements is decided at the moment that the participant arrives by the combination of the per-comment state and the moderation scheme that determines the visibility of unmoderated statements. 

Participants get no feedback about whether their comment was moderated in or out.

\subsection{Seeding}
\label{sec:seeding}
\begin{figure}[htbp]\centering
\includegraphics{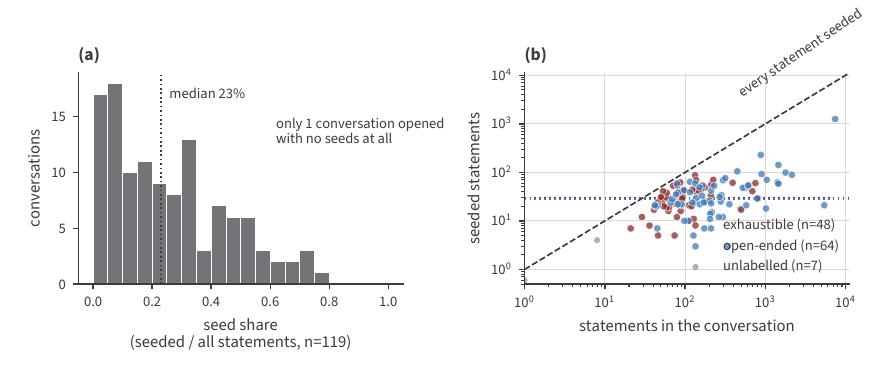}
\caption{What conveners prepared, over the \prov{119} analysed conversations. The share of a
conversation's statements that its convener seeded varies far more than the number they wrote.}
\label{fig:seed-share}
\end{figure}

The convener is expected to seed the conversation with a number of statements. Throughout this section a seed is a statement written by the convener before the first vote cast by anyone other than a statement's own author; statements the convener adds once the conversation is running are treated separately in Section~\ref{sec:mid-conversation-seeding}. On that definition, conveners in the \prov{119} analysed conversations prepared a median of \prov{23} statements and a mean of \prov{25}, with an interquartile range of \prov{16} to \prov{32} and a maximum of \prov{142}. Figure~\ref{fig:seed-share} shows the distribution; all but one conversation carries at least one seed.

Figure~\ref{fig:seed-share} also shows how much of each conversation's final statement set the convener had prepared in advance. The median conversation closes with \prov{17.6}\% of its eventual statements already in place as seeded from the beginning, with an interquartile range of \prov{4} to \prov{32}\% and a maximum of \prov{77}\%. No conversation was entirely seeded.

Out of the \prov{2{,}805} opening seeds that drew at least seven votes from participants other than their author, \prov{1{,}965} (\prov{70.1}\% ) found a majority of respondents agreeing. That is a majority, but it is the lowest rate in the corpus: \prov{84.2}\% of statements written by participants (approved and receiving 7+ votes), and \prov{81.6}\% of statements the convener added once the conversation was running reach majority agreement. At the level of individual votes the gap is sharper, and does not depend on how passes are counted: \prov{29.4}\% of votes cast on an opening seed are disagreements, against \prov{17.0}\% for a participant statement. 

If this high percentage is primarily caused by the asymmetric routing priorities, these statistics should be different based on the routing policy experienced in the 2025--2026 era.

Whether this should be interpreted as the conveners being more willing to frame a statement that the participants can agree to, or as participants feeling pressured by the high number of agreeable statements to make additional statements in a framing that the majority would agree to this, cannot be determined immediately. However, this is where the 2025/2026 software error may provide an interesting natural experiment. If it is possible to determine with high confidence which conversations were subject to the uniform prioritization, this could be leveraged to understand the effect of asymmetric routing is.

\subsection{Statement moderation}
\label{sec:strict-moderation}
\prov{59} of the \prov{119} conversations analysed here are recorded as strictly moderated. A policy is recorded for all but \prov{4} of them, so among those with a known policy the share is \prov{59} of \prov{115} (\prov{51}\%). Across the wider corpus of \prov{271} it is lower: \prov{95} of the \prov{255} with a known policy, or \prov{37}\%.

There is every indication that conveners of the conversations in this dataset did not change this setting mid-conversation, but that would be possible as well and could affect the dynamics as the setting is applied at the time of participation to the statements that are neither explicitly accepted nor rejected. 

The setting is only part of the story: it determines the default approach to submitted statements. When statements are submitted, they carry the status `unmoderated', after which they can be approved or rejected. The strictness only determines whether unmoderated statements are shown or not. Whether the default policy is strict or permissive moderation, the moderator still has the ability to accept or reject each individual statement. 

The acceptance or rejection of a statement is not logged explicitly, and can only be derived from the final state and whether the statement received any votes. Thanks to the prioritization it is fairly safe to assume that the first non-author vote follows shortly after submission. If the statement is rejected after having been votable for a while and receiving votes, these statements and their votes will be `zeroed out' in the clustering matrix and ignored. These votes are however still part of the exported datafiles. 

In total, the corpus contains \prov{180{,}944} votes on subsequently rejected statements. \prov{11}\% of these were automatic votes by the authors, but \prov{89}\% were `real' votes. When the rejected statements are analyzed more closely, it turns out that they are significantly more met with disengagement (`pass') than disagreement. Comparing the \prov{1{,}212} rejected statements with at least ten votes from other participants against statements in the same conversation at the same vote count, the rejected statements drew \prov{9.3} points less agreement (95\% CI \prov{4.4} to \prov{14.1}), but not significantly more disagreement (\prov{+2.6} points, CI \prov{-1.0} to \prov{+6.4}); the deficit is made up mostly by passes (\prov{+6.7} points, CI \prov{+3.8} to \prov{+9.8}). 

Section~\ref{sec:distributions-two-kinds} showed that there are both exhaustible and open-ended conversations. Establishing that first required estimating how many statements a participant was able to vote on during their session. In the former, submitted statements may still be considered by the convener after the process, but will never be seen by other participants. This is however not necessarily an intended consequence by the convener: if a lot of the participants arrive before the convener accepts/rejects statements, that can affect to a large extent how many people get to see these newly submitted statements. Exhaustible is therefore not an intent, but a descriptive term: participants tended to vote to most of the statements available to them.
In Table~\ref{tab:retention} the approval rates are split out for each type of conversation. 

\begin{table}[htbp]
\centering
\footnotesize
\caption{What happens to participant-submitted statements, by kind of conversation. Seeds are excluded: a convener does not reject their own. \emph{Pooled} counts every statement equally; \emph{median} counts every conversation equally. The two rank the regimes in opposite directions.}
\label{tab:retention}
\begin{tabular}{lrr}
\toprule
 & \textbf{Exhaustible} & \textbf{Open-ended} \\
\midrule
Conversations & 48 & 64 \\
Statements submitted & 5,418 & 31,948 \\
Approved & 1,825 & 12,375 \\
Rejected & 2,616 & 14,310 \\
Neither (state 0) & 977 & 5,263 \\
\midrule
Rejected, pooled over statements & 48.3\% & 44.8\% \\
Rejected, median over conversations & 14.0\% & 30.1\% \\
\midrule
Seen by another participant, pooled & 40.4\% & 45.8\% \\
Seen by another participant, median & 89.8\% & 87.9\% \\
\bottomrule
\end{tabular}
\end{table}

\begin{figure}[htbp]\centering
\includegraphics{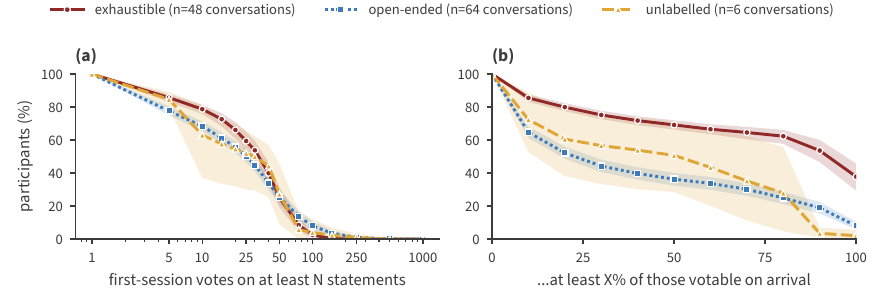}
\caption{The share of the available statements a participant responded to in their first session,
by kind of conversation. The mass at the right-hand edge is participants who responded to
everything available to them.}
\label{fig:coverage-distribution}
\end{figure}

The median exhaustible conversation carries \prov{85} statements against \prov{208} in the median open-ended one. However, the median participant in their first session cast \prov{33} votes against \prov{27}, and spent the same time doing it: \prov{226} seconds in both, with \prov{8.4}\% and \prov{8.3}\% returning for a second session. Participants faced with fewer statements voted more. 

This turns out to be related to the number of statements that was available to the participant at the moment that they were participating. In the exhaustible conversations the median participant completed  \prov{93}\% of the statements available to them, against \prov{15}\% in open-ended ones. 

While it is hard to exactly establish from the available data how many statements were available, by combining multiple signals, that calculation rests on an estimate of what was votable at the moment each participant arrived, and for \prov{4}\% of exhaustible participants the votes cast exceed it, so it should be read as an indication rather than a rate. 

Comparing the distribution of first-session participants (Figure~\ref{fig:coverage-distribution}) shows that the first sessions in exhaustible conversations respond to a higher percentage of statements than those under an open-ended regime. This suggests a hypothesis that participants who are almost done with their assigned set may try to complete it. This may be a finding that platform designers could leverage to design statement stacks of completable size, to encourage participants to complete them. 

\subsection{Mid-conversation seeding}
\label{sec:mid-conversation-seeding}
Section~\ref{sec:seeding} discusses how conveners can seed statements at the start of the conversation. However, this option is also available once the conversation has already started. In \prov{61}\% of the conversations the convener uses this possibility, adding a median of \prov{9} further statements --- a mean of \prov{34}, inflated by one conversation that added \prov{1{,}247}. On average a conversation agreed with \prov{75.6}\% of the statements its convener added mid-conversation, against the \prov{81.6}\% pooled across all such statements reported above. The two are the same computation under two weightings, and the difference of \prov{6.0} points is not resolved by this corpus (95\% CI \prov{-3.0} to \prov{+11.8}, conversations resampled jointly): one conversation contributes half the pooled base and \prov{1} of \prov{71} conversations to the unweighted mean. The rate moves by under two points as the vote screen is varied from five to twenty, and excluding that one conversation moves the pooled rate to \prov{77.1}\%.

A convener who wants to adjust a statement can't edit it. The only way to fix a typo or reword a proposal is to reject the original and submit a replacement as a new seed, and the export records neither of them. The corpus does carry the trace such a workflow would leave. Comparing each mid-conversation seed against the participant statements written before it, and against an equal number written after it (same conversation, same topic, same language, matched for the size of the candidate pool) seeds resemble what preceded them far more than what followed: \prov{9.8}\% against \prov{0.2}\% at a near-duplicate threshold, a ratio of roughly \prov{40}. The comparison is directional by construction, since a seed cannot reproduce a statement that did not yet exist when it was written. The direction holds within conversations as well as across them, in \prov{47} of \prov{70}, and a conversation-clustered interval on the paired rate excludes chance.

When it appears, it is one statement at a time --- \prov{98.4}\% of matched seeds resemble exactly one earlier statement, and none more than two at the same threshold. This method identified the pattern at least once in \prov{26} of the \prov{70} conversations that seed mid-run, and the median such conversation shows it in none of its seeds; there is no evidence of conveners consolidating sets of statements this way. Half of the eligible seeds sit in a single large conversation, which shows the pattern less than the rest, so the rate is not carried by it. Nothing in the export confirms that any particular pair is a replacement rather than two people expressing the same idea; what the comparison establishes is a direction across the corpus, not any individual case.

\section{The participant's choices}
\label{sec:participant-choices}
The baseline Polis is not optimized for authenticating users and making sure they can easily log in again for a second session. The platform is optimized for ease of access, not ease of identification\citep{compdem_identity}. The consequence is that identity is not always tracked very well over time.

The definitions used here:

\begin{itemize}
  \item Participant: A series of votes submitted with the same identification cookie. This usually means same device and same person - but someone who clears or doesn't store cookies may count as multiple people
  \item Session: A series of votes by the same participant, separated from the next vote (if any) by at least an hour.
  \item Vote: the expression of an opinion on a single statement
  \item Statement submission: a comment or statement submitted by a participant.
\end{itemize}

\subsection{Voting pattern}
\label{sec:voting-pattern}
Table~\ref{tab:voting-pattern} displays the distribution of first-session behaviour, pooled over
the \prov{113{,}257} participants in the \prov{118} conversations carrying participant-level data.

\begin{table}[htbp]
\centering
\footnotesize
\setlength{\tabcolsep}{5pt}
\caption{Participant voting patterns over the analysis cohort: \prov{118} conversations, \prov{113,257} participants. Percentiles are taken over participants pooled across conversations. Vote shares include the forced agree Polis injects on a participant's own statement, matching Table~\ref{tab:participation}; only the \prov{15.8}\% who authored anything are affected.}
\label{tab:voting-pattern}
\begin{tabular}{lrrrrr}
\toprule
 & \textbf{10th} & \textbf{25th} & \textbf{50th} & \textbf{75th} & \textbf{90th} \\
\midrule
Votes in first session       & 1 & 6 & 28 & 63 & 136 \\
Time in first session (s)    & 0 & 49 & 239 & 670 & 1467 \\
Agree \%                     & 21.1 & 46.4 & 61.5 & 78.6 & 100.0 \\
Disagree \%                  & 0.0 & 2.6 & 17.3 & 33.3 & 50.0 \\
Pass \%                      & 0.0 & 0.0 & 9.5 & 24.5 & 47.4 \\
Statements submitted         & 0 & 0 & 0 & 0 & 1 \\
\bottomrule
\end{tabular}
\end{table}

Across all conversations, participants have an elasticity of \textbf{0.190} \texttt{[0.122, 0.264]}: a ten percent statements would result in about two percent more votes per participant. The vast majority of participants only participate in a single session of votes, which lasts on average about 3 minutes. This may be an artifact of poor authentication, and I cannot rule out that returning participants are recorded as new ones. 

\begin{table}[htbp]
\centering
\footnotesize
\setlength{\tabcolsep}{4pt}
\caption{Participation by conversation regime, over the analysis cohort. Medians with 95\%
confidence intervals. \emph{All} is the 118 conversations carrying participant-level data;
48 are exhaustible and 64 open-ended, the remaining 6 carrying no regime label.}
\label{tab:participation}
\begin{tabular}{lccc}
\toprule
 & \textbf{All (118)} & \textbf{Exhaustible (48)} & \textbf{Open-ended (64)} \\
\midrule
Participants per conv.        & 316.5 [240.0, 397.5] & 203.0 [166.0, 265.0] & 472.5 [384.9, 561.0] \\
Sessions per participant      & 1.0 [1.0, 1.0] & 1.0 [1.0, 1.0] & 1.0 [1.0, 1.0] \\
Single-session share          & 89.1\% [82.0, 93.8] & 91.6\% [89.0, 93.9] & 91.7\% [89.8, 94.8] \\
First-session duration (s)    & 239 [188, 281] & 236 [177, 286] & 230 [174, 275] \\
First-session votes           & 30.0 [25.0, 32.5] & 33.8 [30.2, 41.0] & 23.5 [20.0, 28.0] \\
\bottomrule
\end{tabular}
\end{table}

Participants tend to spend slightly more time on a statement where they disagree with most of the others, than when they agree with them. A participant takes \prov{10.1}\% longer for a statement when they ddisagree with the majority than when they agree (95\% CI\prov{+9.3} to \prov{+11.1}\%, resampling whole conversations), with similar statement controversy and length. This direction holds in \prov{153} of the \prov{155} conversations measured individually. 
In the same vein, \prov{24.9}\% of minority-side votes take longer than ten seconds against \prov{18.1}\% of majority-side ones. The excess is limited to brief hesitation: the difference disappears at 60s. These
figures are computed over \prov{155} conversations rather than over the \prov{119}-conversation cohort used elsewhere in this paper.

\subsection{Returning participants}
\label{sec:returning-participants}
The number of participants that demonstrably returns for a second session in the same conversation (with a gap of \textbf{at least 1 hour} (following \cite{halfaker2015user})between the last vote of their first session and the first vote of their last session) is on average 8.5\%. The median gap is 15\,h. There is a relatively high number of returning sessions at 1-3 h, which could arguably be considered an extended break rather than a new session.

\begin{figure}[htbp]\centering
\includegraphics{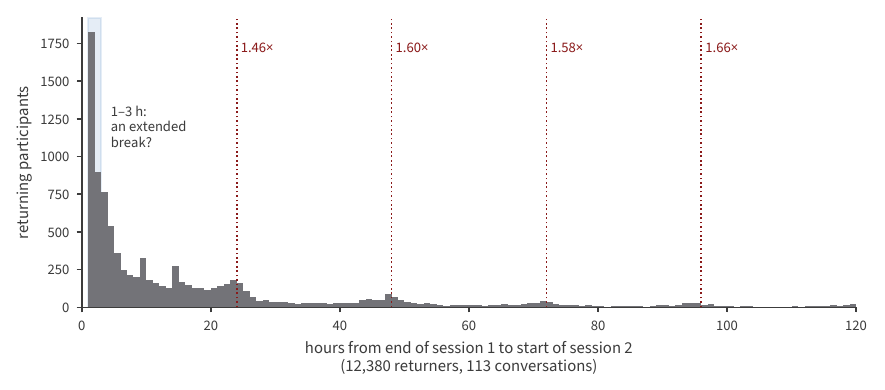}
\caption{The gap between a participant's first and second session, pooled over \prov{12{,}380} returning participants in \prov{113} conversations. Returns concentrate at multiples of twenty-four hours, at roughly one and a half times the surrounding rate.}
\label{fig:return-clock}
\end{figure}

Figure~\ref{fig:return-clock} shows the gap between a participant's first and second session, and identifies clusters at multiples of twenty-four hours -- \prov{1.46}, \prov{1.60}, \prov{1.58} and \prov{1.66} times the surrounding rate at one, two, three and four days. Within a conversation, voting activity is itself concentrated by hour of day, running from about \prov{8}\% of an hour's votes at the busiest hour down to \prov{1}\% twelve hours away. A participant who returns at the same time of day as their first visit therefore produces a gap that is a multiple of twenty-four hours by construction, which is sufficient to account for the peaks without any prompting.

\prov{8.5}\% of participants returns for a second session,

\section{Interpreting Polis Cluster Outputs}
\label{sec:outputs}
One of the exported data files is the clustering of the data as it stood at export time. That clustering cannot be reproduced from the export alone. As described in Section~\ref{sec:polis-produces-grouping}, and documented by the platform's own authors \citep{small2021polis}, each tick begins from the state the previous tick left behind, and the export preserves only the final state — not the intermediate ones, nor the times at which they were computed. Working from the exported data, and against a codebase in which I encountered several bugs, I was unable to recover the reported clusters in any meaningful way.

It also means that the reported clustering is not necessarily the best representation of the final set of opinions: it is where the path happened to arrive, rather than the fit the final votes alone would select.

The number of reported groups is bounded as well: the engine chooses from a small set of candidates, never fewer than two and never more than five (Section~\ref{sec:developer-choices}). In every conversation analysed here the upper bound is five, so the only constraint that ever binds is the floor of two.

If the convener would choose to cluster their independently, there are a few practical details to consider when working with this data: 
\begin{itemize}
    \item The data as considered for clustering is not the same as the data as can be exported. During the clustering process, all meta-statements and rejected (and for strict processes: unmoderated) statements are zeroed out. \item In the database, 'agree' is encoded as -1, and 'disagree' as +1. However, when the data is exported, this is reversed. The way that the container gains access to the data therefore matters how the data should be interpreted. 
    \item Only participants with at least 7 votes or responding to 100\% of the statements at some point during the process are considered in the results. An exception is made for very small conversations. 
\end{itemize}

\section{Discussion and limitations}
\label{sec:limitations}  
The analysis of data from these processes has not only provided an opportunity to learn from the data, but also to identify some unexpected behaviors of the source code. The fact that the source code is open is a blessing here, because it allows researchers to investigate and reproduce what exactly happened. For example, this allowed me to see transparently the bug that changed the priority routing from vote-ratio based to semi-uniform.

In this section, I want to take the opportunity to discuss some recommendations to practitioners (conveners) as well as democratic innovators (platform designers). 

\subsection{Recommendations to practitioners}
\label{sec:practitioners}
I would translate the findings in this paper into a few specific practices that a convener might find helpful without changing their choice of platform. 

First of all, the admission of new statements is an essential part of how Polis works. The policy on which statements to admit and how should be made consciously and preferably in advance, and documented. A conversation that offers a fixed set of statements seems to result in different participant behavior than one where all statements are automatically accepted. While the causality is still unconfirmed, it seems plausible. 

Repeat participation may be limited to a small group of participants, but it is not clear if this is the ceiling or the floor of what's possible. It might be helpful to think about providing a more robust login option to better track participants and allow the convener to actually connect with them. If this is well-designed, a combination with a smaller fixed set at the start may be worth considering. 

Finally, the convener should be aware of the design choices with regards to the clustering and might want to consider to do some additional analysis after the complete set of statements and votes is collected and the process is finalized. At that point, more conscious choices with regards to clustering and what they want to learn from it, can be determined and accounted for.

\subsection{Recommendations to platform designers}
\label{sec:platform-designers}
Platform designers in the \Polispp{} family may also extract some recommendations from this work. 
\begin{itemize}
    \item Recording how a platform behaves is important to be able to reproduce the outcomes, and for researchers to understand the behavior of participants after the fact. For example, what information participants are presented with, or what the probabilities were at the time of assignment. This may help in verifying the fairness of the process, as well as the incremental improvements. 
    \item Specifically to the \Polispp{} family, tracking the statements' lineage might be very helpful to improve the clustering mechanisms. Initial simulations suggest that the result of the clustering can change quite a bit when this is taken into account. In addition, it may be valuable structured information to the convener. 
    \item An independent record of the settings during the process, rather than only at the end, would remove uncertainty about last minute changes. 
\end{itemize}

\subsection{Limitations}
\label{sec:limitations-data}
The dataset also comes with real limitations: because the data collection is not designed as a randomized controlled trial, it is hard to assign causality to the observations. As far as the behavior of conveners is concerned, several of these conversations are organized by the same conveners, or by likely closely connected ones. That does not make the findings less relevant it means the spectrum of possibilities is more informative than the exact percentages.

The same is true for the behavior of participants. The signals found here are valuable stepping stones for future research. However, without a controlled trial, under laboratory conditions or by implementing them in the software, it will be hard to state conclusively whether participants are truly willing to return when more statements are available, or whether dosing statements improves engagement. Nor should it be assumed that all sets of participants would behave the same way: the same interface can be received very differently in one culture than in another.

\section{Conclusion}
\label{sec:conclusion}

The \prov{119} conversations' clusters reflected choices made during the design of the platform as much as by the participants during the conversation. It is impossible to reproduce the exact cluster assignments as reported. Both the warm path choice and the hard limitations on clustering affect even the number of clusters reported, let alone the assignments. While one clustering is not per se better than the other, conveners should do their own due diligence on clustering with complete data at the end of the process. 

From the perspective of a participant, there are two conversation regimes: exhaustible and open-ended. When participants are observing a smaller number of statements that they can respond to, they seem to more frequently complete that entire set of statements, and arrive at a higher median of \prov{33} versus \prov{27} votes. The median participant in an exhaustable conversation completed \prov{93}\% of the statements available to them, while in an open-ended one they completed \prov{15}\%. They do not spend more time doing so, and the effort they put in is the same. In those conversations, about a third of participants completed their entire votable set. This suggests an effect from the achievable goal, but this requires further work to connect causally. 

A median convener seeds their conversation with \prov{23} statements before opening; on average \prov{70.7}\% of a conversation's opening seeds are agreed upon by a majority of the participants that voted on them.

Much of what this paper could not determine traces to a single absence: the exports record what happened, but almost never when. Exports that carried the timing of moderation, of recomputation, and of what each participant was shown would turn most of the open questions here into measurements.

\section*{Declarations}

\subsection*{Declaration of generative AI and AI-assisted technologies in the manuscript preparation process}
During the preparation of this work I used contemporary (2026) Claude Code and supporting LLM tools accessed via API to interactively assist in writing and running analysis scripts, generating tables and figures, and drafting summary prose. I selected all methods, remained in intellectual command throughout, and personally verified the entire analysis pipeline, which is human-reproducible and generates all presented statistics and visualizations. I used these tools to critique and stress-test my own and AI-assisted work; where checks revealed problems, results were narrowed or withdrawn. I do not treat AI-assisted cross-checking as an independent validity guarantee—final judgment and accountability rest with me. No AI tool is an author of this work

\subsection*{Data availability}
A dataset will be made available at the final publication.

\iffinaldraft\else\nocite{*}\fi
\bibliographystyle{plainnat}
\bibliography{refs}

\end{document}